\documentclass[12pt]{article}
\usepackage{times}
\usepackage[super]{natbib}
\usepackage{epsfig, amssymb, amsmath, multicol}
\usepackage{color}
\usepackage{wrapfig} 
\usepackage{hyperref} 
\usepackage[layout=modern]{advancedcoverpage}

\newcommand{\tightitems}{
  \setlength{\topsep}{0pt}
  \setlength{\itemsep}{6pt}
  \setlength{\parsep}{0pt}
  \setlength{\parskip}{0pt}
}

\title{Catching Element Formation In The Act}
\subtitle{The Case for a New MeV Gamma-Ray Mission: \\ Radionuclide Astronomy in the 2020s}
\subsubtitle{A White Paper for the 2020 Decadal Survey}

\author{\emph{\underline{Authors}}\\
Chris L. Fryer, Los Alamos National Laboratory, \href{mailto:fryer@lanl.gov}{fryer@lanl.gov}, (505) 665-3394‬ \\
Frank Timmes, Arizona State University, \href{mailto:fxtimmes@gmail.com}{fxtimmes@gmail.com}, (480) 965-4274‬ \\
Aimee L. Hungerford, Los Alamos National Laboratory \\
Aaron Couture, Los Alamos National Laboratory \\  
~~ \\
~~ \\
Fred Adams, University of Michigan \\
Wako Aoki, National Astronomical Observatory of Japan \\
Almudena Arcones, Technische Universit\"{a}t Darmstadt \\
David Arnett, University of Arizona\\
Katie Auchettl, DARK, Niels Bohr Institute, University of Copenhagen \\
Melina Avila, Argonne National Laboratory \\
Carles Badenes, University of Pittsburgh\\
Eddie Baron, University of Oklahoma \\
Andreas Bauswein, GSI Helmholtzzentrum f\"{u}r Schwerionenforschung \\
John Beacom, Ohio State University \\
Jeff Blackmon, Louisiana State University \\
St\'{e}phane Blondin, CNRS \& Pontificia Universidad Catolica de Chile \\
Peter Bloser, Los Alamos National Laboratory\\
Steve Boggs, UC San Diego\\
Alan Boss, Carnegie Institution for Science \\
Terri Brandt, NASA Goddard Space Flight Center\\
Eduardo Bravo, Universitat Polit\`{e}cnica de Catalunya \\
Ed Brown, Michigan State University \\
Peter Brown, Texas A\&M University \\
Steve Bruenn, University of Florida Atlantic \\
Carl Budtz-J{\o}rgensen, Technical University of Denmark\\
Eric Burns, NASA Goddard Space Flight Center, Universities Space Research Association\\
Alan Calder, Stony Brook University\\
Regina Caputo, NASA Goddard Space Flight Center\\
Art Champagne, University of North Carolina at Chapel Hill\\
Roger Chevalier, University of Virginia \\
Alessandro Chieffi,  Istituto Nazionale di Astrofisica \\
Kelly Chipps, Oak Ridge National Laboratory\\
David Cinabro, Wayne State University \\
Ondrea Clarkson, University of Victoria \\
Don Clayton, Clemson University \\
Alain Coc, Universit\'{e} Paris\\
Devin Connolly, TRIUMF\\
Charlie Conroy, Harvard University \\
Benoit C\^{o}t\'{e}, Konkoly Observatory \\
Sean Couch, Michigan State University\\
Nicolas Dauphas, University of Chicago\\
Richard James deBoer, University of Notre Dame \\
Catherine Deibel, Louisiana State University \\
Pavel Denisenkov, University of Victoria \\
Steve Desch, Arizona State University\\
Luc Dessart, Universidad de Chile \\
Roland Diehl, Max Planck Institute for Extraterrestrial Physics Garching\\
Carolyn Doherty, Konkoly Observatory \\
Inma Dom\'{i}nguez, University of Granada \\
Subo Dong, Kavli Institute for Astronomy and Astrophysics, Peking University \\
Vikram Dwarkadas, University of Chicago\\
Doreen Fan, Lawrence Berkeley National Laboratory\\
Brian Fields, University of Illinois\\
Carl Fields, Michigan State University\\
Alex Filippenko, University of California Berkeley\\
Robert Fisher, University of Massachusetts Dartmouth \\
Francois Foucart, University of New Hampshire \\
Claes Fransson, Stockholm University \\
Carla Fr\"{o}hlich, North Carolina State University \\
George Fuller, University of California San Diego\\
Brad Gibson, University of Hull \\
Viktoriya Giryanskaya, Princeton University \\
Joachim G\"{o}rres, University of Notre Dame \\
St\'{e}phane Goriely, Universit\'{e} Libre de Bruxelles \\
Sergei Grebenev, Space Research Institute, Russian Academy of Sciences \\
Brian Grefenstette, California Institute of Technology\\
Evan Grohs, Los Alamos National Laboratory\\
James Guillochon, Harvard-Smithsonian Center for Astrophysics \\
Alice Harpole, Stony Brook University\\
Chelsea Harris, Michigan State University\\
J. Austin Harris, Oak Ridge National Laboratory \\
Fiona Harrison, California Institute of Technology\\
Dieter Hartmann, Clemson University\\
Masa-aki Hashimoto, Kyushu University \\
Alexander Heger, Monash University \\
Margarita Hernanz, Institute of Space Sciences \\
Falk Herwig, University of Victoria \\
Raphael Hirschi, Keele University \\
Raphael William Hix, Oak Ridge National Laboratory\\
Peter H\"{o}flich, Florida State University \\
Robert Hoffman, Lawrence Livermore National Laboratory \\
Cole Holcomb, Princeton University \\
Eric Hsiao, Florida State University \\
Christian Iliadis, University of North Carolina at Chapel Hill\\
Agnieszka Janiuk, Center for Theoretical Physics Polish Academy of Sciences\\
Thomas Janka, Max Planck Institute for Astrophysics \\ 
Anders Jerkstrand, Max Planck Institute for Astrophysics \\
Lucas Johns, University of California San Diego\\
Samuel Jones, Los Alamos National Laboratory\\
Jordi Jos\'{e}, Universitat Polit\`{e}cnica de Catalunya \\
Toshitaka Kajino, The University of Tokyo \\
Amanda Karakas, Monash University \\
Platon Karpov, University of California Santa Cruz \\
Dan Kasen, University of California Berkeley\\
Carolyn Kierans, University of California Berkeley\\
Marc Kippen, Los Alamos National Laboratory\\
Oleg Korobkin, Los Alamos National Laboratory\\
Chiaki Kobayashi, University of Hertfordshire \\
Cecilia Kozma, Stockholm House of Science \\
Saha Krot, University of Hawaii \\
Pawan Kumar, University of Texas at Austin \\
Irfan Kuvvetli, Technical University of Denmark\\
Alison Laird, University of York  \\
(John) Martin Laming, Naval Research Laboratory \\
Josefin Larsson, KTH Royal Institute of Technology \\ 
John Lattanzio, Monash University \\
James Lattimer, Stony Brook University\\
Mark Leising, Clemson University\\
Annika Lennarz, TRIUMF\\
Eric Lentz, University of Tennessee\\
Marco Limongi, Istituto Nazionale di Astrofisica\\
Jonas Lippuner, Los Alamos National Laboratory\\
Eli Livne, Racah Institute of Physics, The Hebrew University \\
Nicole Lloyd-Ronning, Los Alamos National Laboratory\\
Richard Longland, North Carolina State University \\
Laura A. Lopez, Ohio State University \\
Maria Lugaro, Konkoly Observatory \\
Alexander Lutovinov, Space Research Institute of the Russian Academy of Science \\
Kristin Madsen, California Institute of Technology\\
Chris Malone, Los Alamos National Laboratory\\
Francesca Matteucci, Trieste University, INAF, INFN \\
Julie McEnery, NASA Goddard Space Flight Center\\
Zach Meisel, Ohio University \\
Bronson Messer, Oak Ridge National Laboratory\\
Brian Metzger, Columbia University \\
Bradley Meyer, Clemson University\\
Georges Meynet, University Of Geneva \\
Anthony Mezzacappa, Oak Ridge National Laboratory, University of Tennessee \\
Jonah Miller, Los Alamos National Laboratory\\
Richard Miller, Johns Hopkins University Applied Physics Laboratory\\
Peter Milne, University of Arizona\\
Wendell Misch, Shanghai Jiao Tong University \\
Lee Mitchell, Naval Research Laboratory\\
Philipp M\"{o}sta, University of California Berkeley \\
Yuko Motizuki, RIKEN Nishina Center \\
Bernhard M\"{u}ller, Monash University \\
Matthew Mumpower, Los Alamos National Laboratory\\
Jeremiah Murphy, Florida State University \\
Shigehiro Nagataki, RIKEN  \\
Ehud Nakar, Tel Aviv University \\
Ken'ichi Nomoto, Tokyo University \\
Peter Nugent, Lawrence Berkeley National Laboratory \\
Filomena Nunes, Michigan State University \\
Brian O'Shea, Michigan State University \\
Uwe Oberlack, Johannes Gutenberg University \\
Steven Pain, Oak Ridge National Laboratory\\
Lucas Parker, Los Alamos National Laboratory\\
Albino Perego, Universit\'{a} degli Studi di Milano Bicocca \\
Marco Pignatari, University of Hull \\
Gabriel Mart\'{i}nez Pinedo, Technische Universit\"{a}t Darmstadt \\
Tomasz Plewa, Florida State University \\
Dovi Poznanski, Tel Aviv University \\
William Priedhorsky, Los Alamos National Laboratory\\
Boris Pritychenko, Brookhaven National Laboratory\\
David Radice, Institute for Advanced Study, Princeton University  \\
Enrico Ramirez-Ruiz, University of California Santa Cruz\\
Thomas Rauscher, University of Basel and University of Hertfordshire \\
Sanjay Reddy, Institute for Nuclear Theory, University of Washington \\
Ernst Rehm, Argonne National Laboratory \\
Rene Reifarth, Goethe Universit\"{a}t Frankfurt  \\
Debra Richman, Michigan State University, National Superconducting Cyclotron Laboratory\\
Paul Ricker, University of Illinois \\
Nabin Rijal, Florida State University, National Superconducting Cyclotron Laboratory\\
Luke Roberts, Michigan State University, National Superconducting Cyclotron Laboratory\\
Friedrich R\"{o}pke, Universit\"{a}t Heidelberg, Heidelberg Institute for Theoretical Studies \\
Stephan Rosswog, Stockholm University \\
Ashley J. Ruiter, University of New South Wales Canberra \\
Chris Ruiz, TRIUMF\\
Daniel Wolf Savin, Columbia University \\
Hendrik Schatz, Michigan State University\\
Dieter Schneider, Los Alamos National Laboratory\\
Josiah Schwab,  University of California Santa Cruz\\
Ivo Seitenzahl, University of New South Wales Canberra\\
Ken Shen, University of California Berkeley\\
Thomas Siegert, Max Planck Institute for Extraterrestrial Physics Garching \\
Stuart Sim, Queen's University Belfast \\
David Smith, University of California Santa Cruz\\
Karl Smith, Los Alamos National Laboratory\\
Michael Smith, Oak Ridge National Laboratory \\
Jesper Sollerman, The Oskar Klein Centre, Department of Astronomy \\
Trevor Sprouse, University of Notre Dame\\
Artemis Spyrou, Michigan State University \\
Sumner Starrfield, Arizona State University\\
Andrew Steiner, University of Tennessee, Knoxville, Oak Ridge National Laboratory\\
Andrew W. Strong, Max Planck Institut f\"{u}r Extraterrestrische Physik \\
Tuguldur Sukhbold, Ohio State University\\
Nick Suntzeff, Texas A\&M University\\
Rebecca Surman, University of Notre Dame\\
Toru Tanimori, Kyoto University\\
Lih-Sin The, Clemson University\\
Friedrich-Karl Thielemann, University of Basel and GSI Darmstadt \\
Alexey Tolstov, Open University of Japan, University of Tokyo \\
Nozomu Tominaga, Konan University \\
John Tomsick, University of California Berkeley \\
Dean Townsley, University of Alabama \\
Pelagia Tsintari, Central Michigan University\\
Sergey Tsygankov, University of Turku \\
David Vartanyan, Princeton University\\
Tonia Venters, NASA Goddard Space Flight Center \\
Tom Vestrand, Los Alamos National Laboratory\\
Jacco Vink, University of Amsterdam \\
Roni Waldman, Hebrew University \\
Lifang Wang, Texas A\&M University \\
Xilu Wang, University of Notre Dame\\
MacKenzie Warren, Michigan State University\\
Christopher West, Concordia University \\
J. Craig Wheeler, University of Texas at Austin \\
Michael Wiescher, University of Notre Dame\\
Christoph Winkler, European Space Agency \\
Lisa Winter, Los Alamos National Laboratory\\
Bill Wolf, Arizona State University\\
Richard Woolf, Naval Research Laboratory\\
Stan Woosley, University of California Santa Cruz\\
Jin Wu, Argonne National Laboratory\\
Chris Wrede, Michigan State University\\
Shoichi Yamada, Waseda University \\
Patrick Young, Arizona State University\\
Remco Zegers, Michigan State University \\
Michael Zingale, Stony Brook University \\
Simon Portegies Zwart, Leiden University
}

\presentedto{\underline{Thematic Areas:} \\
{\it PRIMARY: Stars and Stellar Evolution} \\
{\it SECONDARY: Galaxy Evolution} \\
}

\organization{\underline{Projects/Programs Emphasized:} \\
1. All-sky Medium Energy Gamma-ray Observatory (AMEGO)\\
\href{https://asd.gsfc.nasa.gov/amego/}{\tt https://asd.gsfc.nasa.gov/amego/}\\
2. e-ASTROGAM \\
\href{http://eastrogam.iaps.inaf.it}{\tt http://eastrogam.iaps.inaf.it}\\
3. Compton Spectrometer and Imager (COSI) \\
\href{http://cosi.ssl.berkeley.edu}{\tt http://cosi.ssl.berkeley.edu} \\
4. Electron-Tracking Compton Camera (ETCC) \\
5. Lunar Occultation Explorer (LOX)  \\
}
\date{}

\begin{document}
\maketitle

\vspace{-0.13cm}
\section{Executive Summary}\label{sec:exec}
\vspace{-0.2cm}

Gamma-ray astronomy explores the most energetic photons in nature 
to address some of the most pressing puzzles in contemporary
astrophysics.  It encompasses a wide range of objects and phenomena:
stars, supernovae, novae, neutron stars, stellar-mass black holes,
nucleosynthesis, the interstellar medium, cosmic rays and
relativistic-particle acceleration, and the evolution of galaxies. 
MeV $\gamma$-rays provide a unique probe of nuclear
processes in astronomy, directly measuring radioactive decay, nuclear
de-excitation, and positron annihilation.  The substantial information
carried by $\gamma$-ray photons allows us to see deeper into these
objects, the bulk of the power is often emitted at $\gamma$-ray
energies, and radioactivity provides a natural physical
clock that adds unique information.

\vspace{0.1in}
New science will be driven by time-domain population studies at
$\gamma$-ray energies. This science is enabled by next-generation
$\gamma$-ray instruments with one to two orders of magnitude better
sensitivity, larger sky coverage, and faster cadence than all previous
$\gamma$-ray instruments.  This transformative capability permits: (a)
the accurate identification of the $\gamma$-ray emitting objects and
correlations with observations taken at other wavelengths and with
other messengers; (b) construction of new $\gamma$-ray maps of the
Milky Way and other nearby galaxies where extended regions are
distinguished from point sources; and (c) considerable serendipitous
science of scarce events -- nearby neutron star mergers, for example. 
Advances in technology push the performance of new $\gamma$-ray 
instruments to address:
\begin{itemize}\tightitems

\item[$\star$] How do white dwarfs explode as Type Ia Supernovae (SNIa)?

\item[$\star$] What is the distribution of $^{56}$Ni production within a large population of SNIa?

\item[$\star$] How do SNIa $\gamma$-ray light curves and spectra correlate with their UV/optical/IR counterparts?

\item[$\star$] How do massive stars explode as core-collapse supernovae?

\item[$\star$] How are newly synthesized elements spread out within the Milky Way Galaxy?

\item[$\star$] How do the masses, spins, and radii of compact stellar remnants result from stellar evolution?

\item[$\star$] How do novae enrich the Galaxy in heavy elements?

\item[$\star$] What is the source that drives the morphology of our Galaxy's positron annihilation $\gamma$-rays?

\item[$\star$] How do neutron star mergers make most of the stable r-process isotopes?

\end{itemize}

Over the next decade, multi-messenger astronomy will probe the rich
astrophysics of transient phenomena in the sky, including light curves
and spectra from supernovae and interacting binaries, gravitational
and electromagnetic signals from the mergers of compact objects, and
neutrinos from the Sun, massive stars, and the cosmos.  During this new era, the
terrestrial Facility for Rare Isotope Beams (FRIB) 
and Argonne Tandem Linac Accelerator System (ATLAS)
will enable
unprecedented precision measurements of reaction rates with novel
direct and indirect techniques to open perspectives on transient
objects such as novae, x-ray bursts, kilonovae, and the rapid neutron
capture process.  This ongoing explosion of activity in
multi-messenger astronomy powers theoretical and computational
developments, in particular the evolution of community-driven,
open-knowledge software instruments.  {\it The unique information
provided by MeV $\gamma$-ray astronomy to help address these
frontiers makes now a compelling time for the astronomy community to
strongly advocate for a new $\gamma$-ray mission to be operational
in the 2020s and beyond.}

\clearpage

\vspace{-0.3cm}
\section{Supernovae And Other Cosmic Explosions}\label{sec:boom}
\vspace{-0.2cm}

\vspace{0.0in}
\begin{wrapfigure}{l}{3.6in}
\vspace{-0.40in}
\begin{center}
\epsfig{file=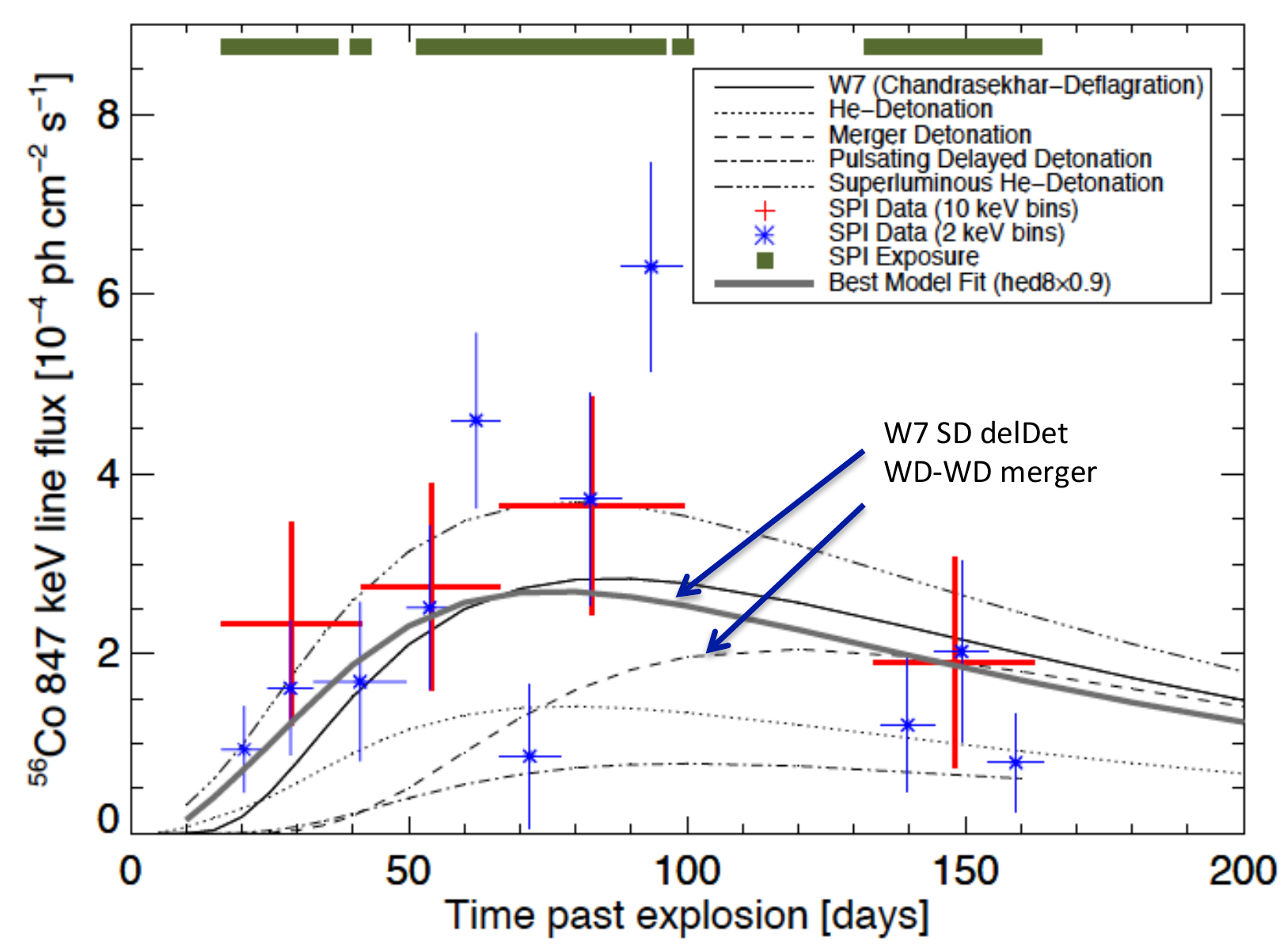,width=3.6in}
\vspace{-0.25in}
\caption{\textit{SN2014J was the first SNIa within reach of current $\gamma$-ray
telescopes\cite{churazov_2014_aa,diehl_2014_aa,diehl_2015_ab}. As the signal from $^{56}$Co
$\gamma$-rays is split into temporal bins, statistical precision is
compromised (blue: 11 time bins; red: 4 time bins; 1D models are shown
as dashed/dotted/solid curves). 
Non-spherical effects may be more important than 1D models indicate,
based on the measurements of radiation processed by the supernova
envelope. A future $\gamma$-ray telescope will measure many SNIa with
a significantly improved precision that complements UV/optical/IR
measurements.
\label{fig:sn2014j}
}}
\end{center}
\vspace{-0.20in}
\end{wrapfigure}

Empirically, SNIa are the most useful, precise,
and mature tools for determining astronomical 
distances\cite{howell_2011_aa}. Acting as standardizable
candles\cite{colgate_1979_aa,phillips_1993_aa,phillips_1999_aa} they
revealed the acceleration of the Universe's 
expansion\cite{riess_1998_aa,perlmutter_1999_aa} and are being used to measure
its properties\cite{wood-vasey_2008_aa,graur_2014_aa,foley_2016_ab,petrushevska_2017_aa}. 
In
stark contrast, the nature of the progenitors and how they explode
remains elusive\cite{wang_2012_aa,maoz_2014_aa}.  The lack
of a physical understanding of the explosion introduces uncertainty in the
extrapolations of the characteristics of SNIa to the distant universe.
In addition, SNIa are expected to be a major source of iron in the chemical
evolution of 
galaxies\cite{chevalier_1976_ab,matteucci_1986_aa,timmes_1995_aa,loewenstein_2006_aa,cote_2017_aa},
cosmic-ray accelerators\cite{edmon_2011_aa, sano_2018_aa}, 
kinetic energy sources in galaxy evolution\cite{simpson_2015_aa,martizzi_2015_aa}, 
and a terminus of interacting binary 
star evolution\cite{iben_1984_aa,whelan_1973_aa,thompson_2011_aa,fang_2018_aa}.
Essentially all SNIa light originates in the nuclear $\gamma$-rays
emitted from the radioactive decay of $^{56}$Ni synthesized in the explosion\cite{colgate_1969_aa},
making their detection the cleanest way to measure the poorly constrained $^{56}$Ni mass.
This bonanza of astrophysical puzzles highlights the need for
a multi-spectral approach
to study such explosions --
extending to the deployment in space of a new and significantly
better $\gamma$-ray telescope.

A line sensitivity 1--2 orders of magnitude better than previous
generation instruments ($\simeq$\,1~$\times$~10$^{-7}$ ph~cm$^{-2}$~s$^{-1}$ 
for broad lines over the 0.05--3.0~MeV range) and a large field
of view ($\gtrsim$\,2.5~sr) will, for the first time, unlock
systematic time-domain SNIa population studies.  High-precision 
measurements of the $^{56}$Ni $\gamma$-ray light curve 
(see Fig.~\ref{fig:sn2014j}) can check and improve the
optical/IR derived luminosity-width relation. Measuring SNIa $\gamma$-ray light curves
beginning within 1 day of the shock breaching the stellar surface and
extending to 100 days, coupled with resolving key radionuclide line
features (not just $^{56}$Co)\cite{seitenzahl_2011_ab} 
in the spectra every 5 days, of 10-100
events/yr out to distance of $\leq$\,100 Mpc, will provide a
significant improvement in our understanding of the SNIa progenitor
system(s) and explosion mechanism(s).  Time-domain characterization of
the emergent SNIa $\gamma$-rays will facilitate the extraction of
physical parameters such as explosion energy, total mass, spatial
distribution of nickel masses\cite{dong_2015_aa}, and ultimately lead to 
the astrophysical modeling and understanding of progenitors and
explosion mechanisms. The relevant $\gamma$-ray light
curves can be extracted from integrated MeV spectra (bolometric),
resolved nuclear lines, or physics-motivated energy bands.  Detection
of several SNIa will distinguish between the models; population
studies involving $\gtrsim$\,100 SNIa will be transformational.

An MeV $\gamma$-ray mission will also act as an early time
monitor/alert system of SNIa in dusty environments like the 
Milky Way plane and nearby starburst regions.
Dust obscuration could delay optical/IR identification of
a SNIa for $\gtrsim$\,2 weeks, but a $\gamma$-ray line detection will be a unique
means of identifying SNIa as early as $\lesssim$\,10 days, 
especially if surface $^{56}$Ni exists as suggested 
by SN2014J $\gamma$-ray observations\cite{piro_2012_aa,diehl_2014_aa,isern_2016_aa}
(see Fig.~\ref{fig:sn2014j}), increasing early detection rates
and maximizing science returns.

A new $\gamma$-ray radionuclide mission is timely: the current \textit{INTEGRAL}
and \textit{NuSTAR} missions are in their late phases.
A new $\gamma$-ray radionuclide mission improved by technological advances made in the past decade
will provide unique data of significant interest across a range of topics
to the broad astronomical community, complementary to the
multi-messenger data also provided by \textit{JWST}, \textit{LSST},
\textit{ALMA}, \textit{TESS},  \textit{fermi}, \textit{TMT},
\textit{GMT}, \textit{SKA}, \textit{Gaia}, \textit{IceCube},
\textit{CTA}, \textit{JUNO}, \textit{FRIB},  \textit{ATLAS}  and \textit{LIGO}.

Cataclysmic variables are semi-detached binary systems consisting of a
white dwarf accreting from a low mass stellar
companion\cite{kraft_1963_aa,giovannelli_2008_aa,shen_2009_ac,ju_2016_aa,oliveira_2017_aa}.
They are progenitors for nova events, with classical novae being the
most optically luminous subclass\cite{jose_2007_ac,starrfield_2016_aa, mroz_2016_aa}.  
Some classes of novae may be the progenitors of a population of
SNIa\cite{hernanz_2008_aa, shafter_2015_aa, soraisam_2015_aa,starrfield_2016_aa, shafter_2017_aa}.  
Two types of MeV $\gamma$-ray
emission are expected from novae: prompt emission from $e^{-}e^{+}$
annihilation with the $e^{+}$ originating from $^{13}$N and $^{18}$F,
and a longer-lasting emission from $^{7}$Be and $^{22}$Na
decays\citep{leising_1987_aa,gomez-gomar_1998_aa}.  The prompt emission has
a $\lesssim$\,1 day duration and appears $\simeq$\,1--2 weeks before
optical maximum, and the longer-lasting
emission persists for $\simeq$\,0.1--3 yr.
Recent UV detections of a
few novae suggest the $^7$Be ejecta mass is larger than current
1D models produce\cite{tajitsu_2015_aa,tajitsu_2016_aa,molaro_2016_aa}.
A next-generation $\gamma$-ray mission as described above
will allow, for the first time, systematic time-domain studies of
novae populations. Such explorations will address key uncertainties
about mixing between the accreted matter and the white dwarf, the
conversion of radioactivity into optical emission, and the
contribution of novae to galactic enrichment. In addition,
measurements at facilities such as
\textit{ARIEL}, \textit{ATLAS}, and \textit{FRIB}
and stable beam facilities
will approach a complete set of reaction rates 
for classical novae\cite{arcones_2017_aa}
on a similar timeline for a next-generation gamma-ray mission.

\vspace{0.0in}
\begin{wrapfigure}{l}{3.3in}
\vspace{-0.10in}
\begin{center}
\epsfig{file=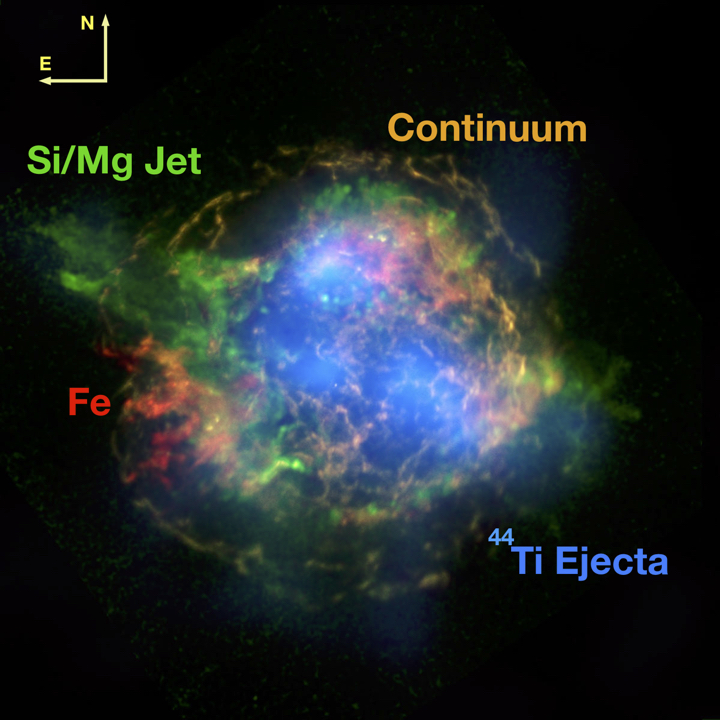,width=3.3in}
\caption{\label{fig:casa} \textit{
3D distribution of Cas~A ejecta.\hfil\ \break
{\it NuSTAR} $^{44}$Ti in blue,
{\it Chandra} continuum in gold, 
Si/Mg band in green,
X-ray emitting iron in red\cite{grefenstette_2017_aa}.
}}
\end{center}
\vspace{-0.20in}
\end{wrapfigure}

Other cosmic explosions such as core-collapse supernovae (CCSN),
pair instability supernovae, neutron star mergers, fast radio bursts, 
and gamma-ray bursts are also expected to exhibit key signatures 
about their interior workings that can be observed with a 
modern \hbox{$\gamma$-ray} telescope. For example,
the spatial distribution of elements in young supernova remnants
directly probes the dynamics and asymmetries traced by, or produced
by, explosive nucleosynthesis\cite{nagataki_1998_aa,wongwathanarat_2017_aa}.

One crucial diagnostic in young remnants
is the relative production of $^{44}$Ti, $^{56}$Ni, and $^{28}$Si. 
These have indirectly been
observed in X-rays from atomic transitions, and $\gamma$-rays from
radioactive decay have shown how this can be misleading 
about where the newly-formed elements actually reside
(see Fig. \ref{fig:casa}).  The physical processes
that produce these isotopes in CCSN depend on the
local conditions of the shock during explosive
nucleosynthesis\cite{the_1998_aa,the_2006_aa,young06,timmes_1996_ab,magkotsios10,chieffi_2017_aa}.  
The isotope $^{44}$Ti
($\tau_{1/2} \simeq$\,60~yr\cite{hashimoto_2001_aa,ahmad_2006_aa}) 
offers a key diagnostic of the explosion 
mechanism\cite{renaud_2006_aa,hwang_2012_aa,grebenev_2012_aa,boggs_2015_aa,grefenstette_2014_aa,grefenstette_2017_aa}
because its synthesis is the most sensitive to
the local conditions.  For example, Cas A was an excellent target for 
current $\gamma$-ray instruments because it is young ($\simeq$\,340\,yr)
and nearby ($\simeq$\,3.4\,kpc).  Its ejecta has been monitored for
decades at X-ray/optical/IR wavelengths, which are now understood to only provide
complementary insight into the dynamics and asymmetries of a 
young supernova remnant\cite{fesen_2001_aa,delaney_2010_aa,rest_2011_aa,milisavljevic_2013_ab,alarie_2014_aa,lopez_2018_aa}, 
while radioactive decay unambiguously traces the flow and dynamics of new ejecta.

To date, only Cas A and SN 1987A have 
been used to place constraints on CCSN progenitors and explosion 
mechanisms\cite{the_1990_aa,shtykovskiy_2005_aa, larsson_2011_aa, seitenzahl_2014_aa, tsygankov_2016_aa}.
A new MeV $\gamma$-ray mission with the characteristics
described above will detect $\simeq$\,8 young supernova remnants in
the Milky Way\cite{dufour_2013_aa} and provide
a precise abundance measurement of $^{44}$Ti in the remnant of SN 1987A\cite{grebenev_2012_aa,boggs_2015_aa,tsygankov_2016_aa}.  
New measurements of a few CCSN in 
their $^{44}$Ti light will add to our knowledge; population
studies with a four-times larger sample size to determine
the variation in $^{44}$Ti yields from CCSN will be
groundbreaking.

\vspace{-0.5cm}
\section{Tracing Chemical Evolution}\label{sec:trace}
\vspace{-0.15cm}

\vspace{0.00in}
\begin{wrapfigure}{l}{3.3in}
\vspace{-0.30in}
\begin{center}
\epsfig{file=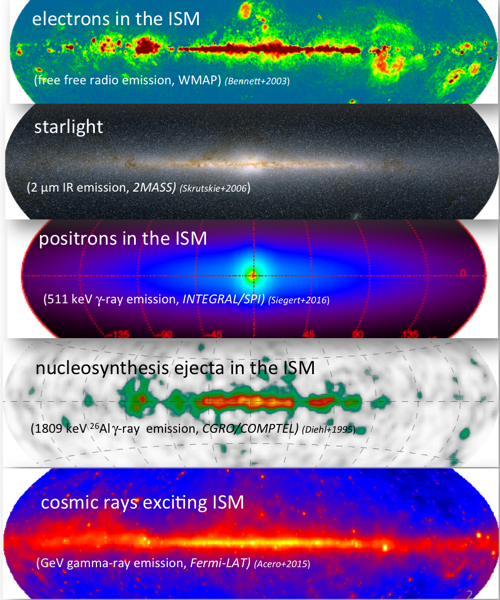,width=3.3in}
\vspace{-0.30in}
\caption{\label{fig:longlived} \textit{Deciphering the Milky Way. A modern MeV $\gamma$-ray instrument 
will help solve how newly created elements are produced, transported, mixed, and distributed. 
}}
\end{center}
\vspace{-0.30in}
\end{wrapfigure}

The star-gas-star cycle operating in the evolution of galaxies
includes at least four phases where MeV $\gamma$-ray astronomy
provides unique and direct diagnostics of cosmic explosions and
chemical evolution. (1) The ejected yields of radionuclides by stars
and explosive nucleosynthesis events tell us about the otherwise
hidden conditions of nuclear fusion reactions in these sites.  (2) The
flow of stellar ejecta into the ambient gas (i.e., mixing in chemical
evolution) is directly traced by radionuclides over their radioactive
lifetimes, which is possible because the $\gamma$-ray emitting nuclear
decays are independent of the thermodynamics or composition
of the ambient gas. (3) Positrons emitted by radioactive decays,
visible through their annihilation $\gamma$-rays, tell us about the
nucleosynthesis in individual events and the structure and dynamics of
the Galaxy.  (4) Nuclear de-excitation $\gamma$-rays caused by cosmic
ray collisions with the ambient gas provide the most direct
measurements of the cosmic ray flux at MeV energies 
and illuminate otherwise invisible fully-ionized gas (e.g., the hot ISM and IGM). 
These four items are the science
drivers for a new $\gamma$-ray mission in the 2020s.


Because their lifetimes are long 
(ten times longer than observables at any other wavelength)
compared to the interval between
massive star supernovae, yet abundant enough to yield
detectable emission when they decay, the radionuclides $^{26}$Al
($\tau_{1/2} \simeq$ 7.3$\times$10$^5$ yr\cite{norris_1983_aa,thomas_1984_aa}) 
and $^{60}$Fe ($\tau_{1/2} \simeq$ 2.6$\times$10$^6$ yr\cite{rugel_2009_aa,ostdiek_2015_aa}) 
are valuable tools of $\gamma$-ray astronomy for advancing our global
understanding of massive stars and their supernova explosions.  This
includes the complex late phases of stellar 
evolution\cite{woosley_2002_aa,burrows_2012_aa,couch_2013_ac,couch_2015_aa,bruenn_2016_aa,muller_2017_aa,oconnor_2018_aa}, 
actions of neutrinos\cite{frohlich_2006_ab,patton_2017_aa,patton_2017_ab} 
and supernova shockwaves\cite{blondin_2003_aa,fernandez_2015_ab}, 
and how ejecta of new elements from these sources are spread in galaxies\cite{kobayashi_2011_aa,vincenzo_2018_aa}.  
The clock inherent to emission from radioactivity again helps here, as in the case of Cas A above, 
to illuminate otherwise invisible, tenuous gas flows.
The short-lived radionuclides $^{26}$Al, $^{60}$Fe, $^{53}$Mn, and $^{182}$Hf present
in the early Solar System play a pivotal role in constraining its
formation and chronology. Furthermore, $^{26}$Al is the major heating source for
thermal and volatile evolution of small planetesimals in the early
Solar System\cite{urey_1955_aa,lee_1976_aa,mostefaoui_2005_aa,tang_2012_aa,lugaro_2018_aa}.

\vspace{0.05 in}
Current $\gamma$-ray instruments measure the diffuse emission from
$^{26}$Al and $^{60}$Fe decays in the inner portions of the Milky Way 
Galaxy\cite{pluschke_2001_aa,smith_2004_aa,diehl_2006_ab,wang_2007_aa} (see
Fig. \ref{fig:longlived}), and the bulk dynamics of $^{26}$Al through
Doppler shifts and broadening of the $\gamma$-ray line for 3--4
massive-star groups / OB associations\cite{martin09,diehl10,kretschmer_2013_aa}.  This
provides a key test for models of stellar feedback in galaxies, including
massive-star winds, supernova explosion energy, and abundance mixing
physics.
A new $\gamma$-ray instrument with a
line sensitivity 1--2 orders of magnitude better than previous
instruments ($\simeq$\,1 $\times$ 10$^{-7}$ ph cm$^{-2}$
s$^{-1}$ for broad lines over 0.05--3.0~MeV), angular resolution of
1--2$^\circ$, and energy resolution of 0.1\% (to
differentiate the emission lines from specific OB associations against the diffuse
radioactive afterglow of stellar activity), 
will increase the number of $\gamma$-ray observed OB associations 
by an order of magnitude to 25--35 based on observed distances to 
OB associations\cite{martin09,diehl10,melnik17}.  

\vspace{0.05 in}
Another signal addressed by the same new $\gamma$-ray telescope is 
positron annihilation and its characteristic $\gamma$-ray spectrum,
including a line at 511~keV. Current telescopes have established a 
morphology of our Galaxy's annihilation $\gamma$-rays peaking in the
inner Galaxy\cite{knodlseder_2005_aa,weidenspointner_2008_aa,siegert_2016_aa}, 
while most candidate sources reside in the Galaxy's disk. 
Solving this puzzle includes re-examining cosmic rays,
supernovae\cite{crocker_2017_aa},
pulsars, microquasars, the Fermi bubbles,
neutron star mergers\cite{fuller_2018_aa},
and possibly dark matter emission.

\vspace{-0.6cm}
\section{When Opportunity Knocks}\label{sec:opp}
\vspace{-0.15cm}

\noindent
A new MeV $\gamma$-ray observatory offers considerable serendipitous
science for uncommon or surprising events such as a nearby
CCSN, neutron star merger, or fast radio burst\cite{delaunay_2016_aa}.   Their detection in
$\gamma$-rays could entirely restructure our understanding of both the
transient itself and its implications for astrophysics as a whole.
For example, a detector with a line sensitivity 50 times
greater than current instruments will detect 7 radioactive isotopes
($^{48}$Cr, $^{48}$V, $^{52}$Mn, $^{56-57}$Co, $^{56-57}$Ni)
from a CCSN occurring within 1\,Mpc and
7 more ($^{43}$K, $^{44}$Ti, $^{44}$Sc, $^{47}$Sc, $^{47}$Ca, $^{51}$Cr, $^{59}$Fe)
if within 50\,kpc.  These radionuclides provide a unique and powerful
probe of the explosion of massive stars\cite{nomoto06,diehl_2018_aa}.  Similarly,
$\gamma$-rays from the radionuclides produced during the
r-process\cite{cote18} in a neutron star merger such as
GW170817\cite{abbott_2017_ad,abbott_2017_af} would
be detectable at 3-10\,Mpc\cite{hotokezaka16}.  Exact
yields from GW170817 are difficult to determine from optical/IR
measurements alone, and it is not settled that GW170817 produced
the heavy r-process elements\cite{lippuner_2017_aa,cote18,holmbeck_2018_aa,horowitz_2018_aa}.
A sufficiently strong $\gamma$-ray signal,
coupled with a set of multi-messenger signals, could
distinguish between light and heavy r-process production to possibly cement
neutron star mergers as the dominant r-process site.

\vspace{-0.6cm}
\section{Imagining the Future}\label{sec:sum}
\vspace{-0.15cm}

\noindent
The time is ripe for the astronomy community to strongly advocate for
a new MeV $\gamma$-ray mission to be operational in the 2020s.
Such a mission will be based on advanced space-proven detector technology
with unprecedented line sensitivity, angular and energy resolution, sky coverage,
polarimetric capability,
and trigger/alert capability for, and in conjunction with, other multi-messenger instruments.
Potential missions include 
\textit{AMEGO}\cite{amego_2018_aa},
\textit{COSI}\cite{kierans_2017_aa},
\textit{e-AstroGAM}\cite{angelis_2018_aa},
\textit{ETCC}\cite{tanimori_2017_aa},
\textit{HEX-P}\cite{madsen_2018_aa},
and \textit{LOX}\cite{miller_2018_aa}.
A new MeV $\gamma$-ray mission will open unique windows on the Universe
by making pioneering observations of cosmic explosions and the flow
of their newly created elements into Galactic ecosystems.

\clearpage

\bibliographystyle{aasjournal}
\begin{multicols}{3}
\begin{scriptsize}
\bibliography{ra2020}
\end{scriptsize}
\end{multicols}

\end{document}